%% file: IEEE-conference.tex
\documentclass[conference]{IEEEtran}
\IEEEoverridecommandlockouts

\usepackage{cite}
\usepackage{amsmath,amssymb,amsfonts}
\usepackage{algorithm}
\usepackage{graphicx}
\usepackage{textcomp}
\usepackage{xcolor}
\usepackage{subfig}
\usepackage{authblk}
\usepackage{algpseudocode}

\def\BibTeX{{\rm B\kern-.05em{\sc i\kern-.025em b}\kern-.08em
    T\kern-.1667em\lower.7ex\hbox{E}\kern-.125emX}}
    
\IEEEoverridecommandlockouts
\IEEEpubid{\makebox[\columnwidth]{ 979-8-3503-6583-
2/25/\$31.00~\copyright~2025~IEEE \hfill}
\hspace{\columnsep}\makebox[\columnwidth]{ }}

\begin{document}

\title{Terahertz Defect Detection in Multi-Layer Materials Using Echo Labeling
\thanks{This work is funded by the German Research Foundation (“Deutsche Forschungsgemeinschaft”) (DFG) under Project–ID 287022738 TRR 196 for Project S02.} }

 

\author{Dogus Can Sevdiren}
\author{Furkan H. Ilgac}
\author{Aydin Sezgin}

\affil{\ Institute for Digital Communication Systems, Ruhr-Universit\"at Bochum, Germany \\  Emails: \{dogus.sevdiren, furkan.ilgac, aydin.sezgin\}@ruhr-uni-bochum.de}

\maketitle

\begin{abstract}
Non-destructive testing is an important technique for detecting defects in multi-layer materials, enabling the evaluation of structural integrity without causing damage on test materials. Terahertz time-domain spectroscopy (THz-TDS) offers unique capabilities for this purpose due to its sensitivity and resolution. Inspired by room geometry estimation methods in acoustic signal processing, this work proposes a novel approach for defect detection in multi-layer composite materials using THz-TDS, enhanced by high-power sources. The proposed method utilizes Euclidean distance matrices to reduce problem complexity compared to state-of-the-art approaches, and effectively distinguishes and maps higher-order reflections from sublayers, enabling precise defect localization in composite materials without artifacts.
\end{abstract}


\section{Introduction}
    \input{1_introduction}
\section{System Model}

\input{2_system_model}
\section{Defect Detection}
    \input{3_defect_detection}  
\section{Results}
    \input{4_results}

\section{Conclusions}
    \input{5_conclusions}

\bibliographystyle{IEEEtran}
\bibliography{references}

\end{document}

%% file: 1_introduction.tex
Terahertz (THz) frequencies (0.3-10 THz) offer unique propagation properties and huge bandwidth, making them attractive for researchers seeking to leverage this spectrum for sensing applications, such as material characterization. An important application of THz sensing is the nondestructive testing, which enables the characterization of material structures without damaging them, offering significant advantages, particularly the internal structure of a material of interest. One of the primary uses of THz material characterization is the detection of internal defects in composite materials, specifically non-impregnated voids \cite{yakoleVoids} and delamination \cite{ospaldDelim}. In this regard, THz time-domain spectroscopy (THz-TDS) emerged as the most prominent method due to its fine time resolution \cite{Lukas}. Especially, THz-TDS systems that employ a single shot technique allow for fast data acquisition with good temporal resolution \cite{TEOSingle}. However, using single-shot techniques require high pulse energy \cite{mansourzadeh2021high}. In this consideration, authors of \cite{Saraceno} developed laser-driven single-cycle high-power emitters enabling the use of single-shot THz-TDS for material characterization with very high spatial resolution. Notably, the generated pulses could penetrate through multiple layers while simultaneously providing strong reflections for detection.
\par

Consider  THz-TDS of a composite multi-layer material under test (MUT) in reflection mode where the pulses are emitted toward the MUT. In this configuration, incident THz pulses interact with the MUT, resulting in reflections from the layers and any present defects within the material structure. Depending on the system geometry, the pulses might initially reflect off the defects, then encounter the boundaries of the sublayers, and finally reflect back towards the detectors. In this case, the defects act as in-material sources, emitting both direct and reflected power toward the receivers. This interplay of reflections creates a scenario analogous to acoustics room shape estimation, where the acoustic waves emitted from sources (speakers) are reflected from the walls \cite{acoustics} \cite{acusticGraph}. In contrast to room-shape estimation, where the locations of the sources are known and walls are of interest, in defect detection problems, the unknowns are the presence, the number, and the number of in-material sources (defects).

In this work, we propose a novel defect detection approach leveraging the analogy with acoustic problems. Specifically, we consider a system that employs a high-power laser source for THz-TDS and utilize echo labeling method based on the Euclidean distance matrices (EDMs) for detecting defects in multi-layer composite materials. 

\subsection{Related Works}

In \cite{yakoleVoids}, the authors employed TDS-based THz tomography to detect defects. In \cite{Sampath}, a method is proposed for the identification of materials within layered objects using morphological component analysis. Furthermore, \cite{SampathMIMO} uses the sparsity of defects for identification through backscattered signals, while \cite{SampathML} applies support vector machines for the same purpose.\par
 To the best of our knowledge, this is the first work to explore defect detection in multi-layer materials using THz-TDS systems using echo labeling.

%% file: 2_system_model.tex
\begin{figure}[t!]
    \centering
    \includegraphics[width=.9\linewidth]{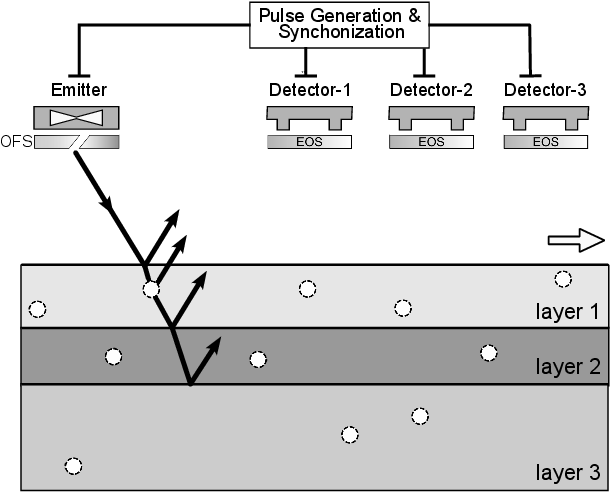}
    \caption{{An illustration of multi-layer defect detection with a TDS system with single emitter (Tx) and multiple  detectors (Rx). OFS: Optical Focusing System. EOS: Electro-Optic Sampling.}}
    \label{fig:system}
\end{figure}

Fig. \ref{fig:system} illustrates the system geometry, showcasing a THz-TDS system equipped with a THz emitter (Tx)
 and multiple receivers (Rx). A terahertz time-domain spectroscopy (THz-TDS) system utilizes ultrafast laser pulses to generate and detect terahertz radiation. The pulse width can be as low as 100 femtoseconds. The system typically employs photoconductive antennas (PCAs) as emitters and electro-optical (EO) techniques for detection  \cite{Mohsen}. In the emitter, a femtosecond laser pulse excites a semiconductor material, creating a transient current, when biased by an external voltage generates a broadband terahertz pulse. This terahertz pulse is directed towards the sample via an optical focusing system (OFS) to interact with its material properties. Typically, for detection, the electro-optical sampling (EOS) technique is used, where the terahertz wave is  demodulated by an electro-optic crystal. A second synchronized laser pulse, known as the probe beam, passes through the crystal, and its polarization is altered in proportion to the terahertz field. By measuring the polarization changes using a balanced photodetector, the temporal profile of the terahertz pulse is reconstructed.

The system is designed to inspect a multi-layer material containing numerous defects distributed across different layers. Given the high directivity of the OFS employed in the emitter, it is assumed that only a narrow part of the material is illuminated with each pulse, while a full inspection of the material can be performed by moving the material across the emitter's boresight. Due to the multiple layers, internal reflections inside the material occur, causing additional echoes reaching to the receivers.

%% file: 3_defect_detection.tex
\begin{figure}[b!]
    \centering
    \includegraphics[width=0.75\linewidth]{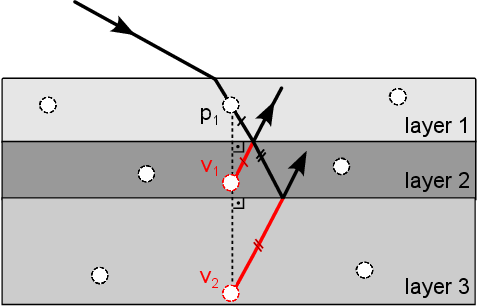}
    \caption{The second and third order reflections due to multi-layer surfaces can be modeled as virtual sources.}
    \label{fig:virtual}
    \vspace{-1mm}
\end{figure}

The pulse excited from the source travels through the material and is subject to reflections in different geometries. Consider a TDS-MUT geometry as given in Fig. \ref{fig:system} top to bottom. The pulse first hits the upper boundary of the material, with some of it reflected, and the rest penetrating to the material. The pulse encountering with the defect inside the material is first reflected by the defect, then hit the boundary of the layer just below it, and  reflect off the lower layers subsequently. Additionally, pulses are also reflected by the sublayer boundaries. Each reflection reaches the receivers on the order of the TOAs, which is proportional to the distance the pulse has traveled, and the receivers have a collection of shifted pulses in time in the respective order.
\par
Having a collection of reflections exposes two problems. First, the received signals contain multiple order reflections from the same defect. This problem can be tackled with so-called image sources. Consider a defect as given in  Fig. \ref{fig:virtual}, the defect first reflects off the pulse directly, but also have secondary and further reflections from the sublayers below. For the first case the geometry reveals the defect, however in the latter, the signal propagates a longer distance and TOAs imply that there are other defects, although this is not true. To mitigate this problem, we can introduce image sources. An image source can be considered as an imaginary defect, and are symmetrical to defects with respect to sublayer boundary. In this case, the distance traveled by the signal reflected from the defect and a sublayer sequentially is the same as the distance from the image source. Exploiting this geometric property, we can reduce the order of reflections in our problem to one, and the first problem reduces to image source localization problem. The second problem is that the receivers might acquire reflections of the defects and image sources sources in a mixed order. This problem can be efficiently solved by echo labeling employing EDMs \cite{acoustics} \cite{acusticGraph}. In the rest of the paper we refer both the defects and the image sources simply as sources.

\subsection{Echo Labeling based on Euclidean Distance Matrices}
The EDM is a square matrix whose entries are the distances between two points. Consider $M$ points, the corresponding EDM of the $M$ points with $\mathbf D \in \mathbb{R}^{M\times M}$. We can write the elements of EDM as $\mathbf D[i,j] = \left|\right| \mathbf p_i - \mathbf  p_j\left|\right|_2^2 $, where $p \in \mathbb R^d$ denotes a $d$ dimensional vector, and $i$ and $j$ denote column indexes, respectively. One important property of EDM is the rank deficiency which we take benefit in echo labeling problem. As given in \cite[Theorem~1]{DokmanicEDM},  the rank of an EDM is independent of the number of points that generate it and is at most $d+2$. Hence the rank of a $2$D-EDM  cannot exceed $4$.
\par
Consider, a THz-TDS system with $N\geq4$ receivers. The EDM of receivers $\mathbf D$ composed of the distances between receivers, will have a rank of $4$. If, one extends this matrix with a vector $\mathbf{d_p}$, which contains the distances of receivers to an arbitrary point $p$, and vice versa, the rank does not change. In other words, if we augment the matrix $\mathbf D$ with $\mathbf{d}_p \in \mathbb{R}^{N \times 1}$ as \cite{acoustics}
\begin{equation}
    \tilde { \mathbf D} = 
    \left[ {\begin{array}{cc}
   \mathbf{D} & \mathbf{d}_p \\
   \mathbf{d}^T_p & 0 \\
  \end{array} } \right],
\end{equation}
where, $p$ indicates a source, the rank of the augmented EDM does not change. Given this information, if one combines the reflections from the same source among receivers, and estimates the perfect distance vector $\mathbf{d}_p$ given perfect TOAs, the rank of $\tilde{\mathbf{D}}$ remains same, i.e. $rank (\tilde{\mathbf{D}}) \leq 4 $, which enables us to find the location the sources within the material, and consequently reconstruct the MUT.
\par
A key drawback of the rank test is their reliance on accurate distance estimations. In practice, TOA measurements are often subject to perturbations, making it difficult to obtain precise estimates on distances. These inaccuracies may lead the augmented matrix $\tilde{\mathbf{D}}$ to be a non-EDM and have a larger rank than $4$. Consequently, excluding all augmented matrices based on rank test would also exclude correct reflection combinations\cite{acusticGraph}. To overcome this issue, we employ using singular value thresholding.
\par 
 \subsection{Echo Labeling in the Presence of Noise}
The rank of a matrix can be interpreted as the number of nonzero singular values. Therefore, by reversing the impact of noise on the number of singular values that are supposed to be equal to zero, we can perform an appropriate rank test. Consider the singular value decomposition of augmented matrix $\tilde{\mathbf{D}}$ as
\begin{equation}
    \tilde{\mathbf{D}} = \sum_{j=1}^{N+1}\lambda_ju_jv_j^T,
\end{equation}
where the singular values of $\tilde{\mathbf{D}}$ denoted by the $\lambda_j$. The thresholding basically sets all singular values less than a certain value, $\sigma$, to zero and can be expressed as  \cite{SHABALIN201367}
\begin{equation}
    G(\tilde{\mathbf{D}})= \sum_{j=1}^{N+1}\lambda_jI(\lambda_j\geq\sigma)u_jv_j^T.
\end{equation}
 where, $\sigma$ depends on the noise variance. Applying thresholding and rank test sequentially, we obtain a set of echo combinations. However, the threshold cannot be set arbitrarily low, the number of elements in the set will be larger than the number of sources. In \cite{acusticGraph}, the authors developed an efficient method to find the right echo combinations given the reduced set based on graph-theory. Basically, the combination set can be expressed in terms of an undirectional graph, the elements of the set are denoted as nodes, and the nodes are connected by an edge if they share a common element, i.e., a shared reflection component. The maximum-independent set of the graph, reveals the correct echo combinations and the cardinality of the maximum independent set is equal to number of sources. While a full exploration of this topic is beyond the scope of this paper, we encourage to read \cite{acusticGraph} for further details. Finding the maximum independent set is a well-studied problem and can be solved via different methods, e.g., Bron-Kerbosch algorithm \cite{BronKerbosch}.

 \subsection{Defect Reconstruction}
 Finding correct echo combination for sources, reveals the distances of the sources to the receivers. Given the position of the receivers and the distances, we can locate the source locations by using least-squares.
 \par
 Given the sources and measurement geometry, we can reconstruct the defects iteratively as given in Alg. \ref{DefRec}. Initially, we identify a potential defect by locating a source from top to bottom. Subsequently, we determine corresponding image sources by exploiting symmetry. To account for measurement noise, we define a circular region of interest with a radius equal to the noise standard deviation, i.e.,  $ r= \sigma $, where the image sources can be present. Within this region, we search for potential image source locations. Once all image sources for a given defect are identified, we apply reverse symmetry to map them around the defect. Finally, we average the positions of all sources associated to same defect to estimate the precise location.

\begin{algorithm}[ht!]
  \caption{Defect Reconstruction}\label{DefRec}
  \begin{algorithmic}[1]
    \State Locate sources using least squares
    \State Define a circular vicinity with a diameter of $r=\sigma$
    \Repeat
    \State Label first source from top as a defect   
    \State Find symmetries of the defect 
    \State \parbox[t]{\dimexpr\linewidth-\algorithmicindent} {Label closest sources within the vicinities of symmetries as image sources of the defect}
    \State \parbox[t]{\dimexpr\linewidth-\algorithmicindent}
    {Take the reverse symmetry of the image sources so that they map around the defect}
    \State Average associated sources to estimate defect location
    \Until{There is no unlabeled sources left}
  \end{algorithmic}
\end{algorithm}

%% file: 4_results.tex
\begin{figure}[b!]
    \vspace{-4mm}
    \centering
    \includegraphics[width=\linewidth]{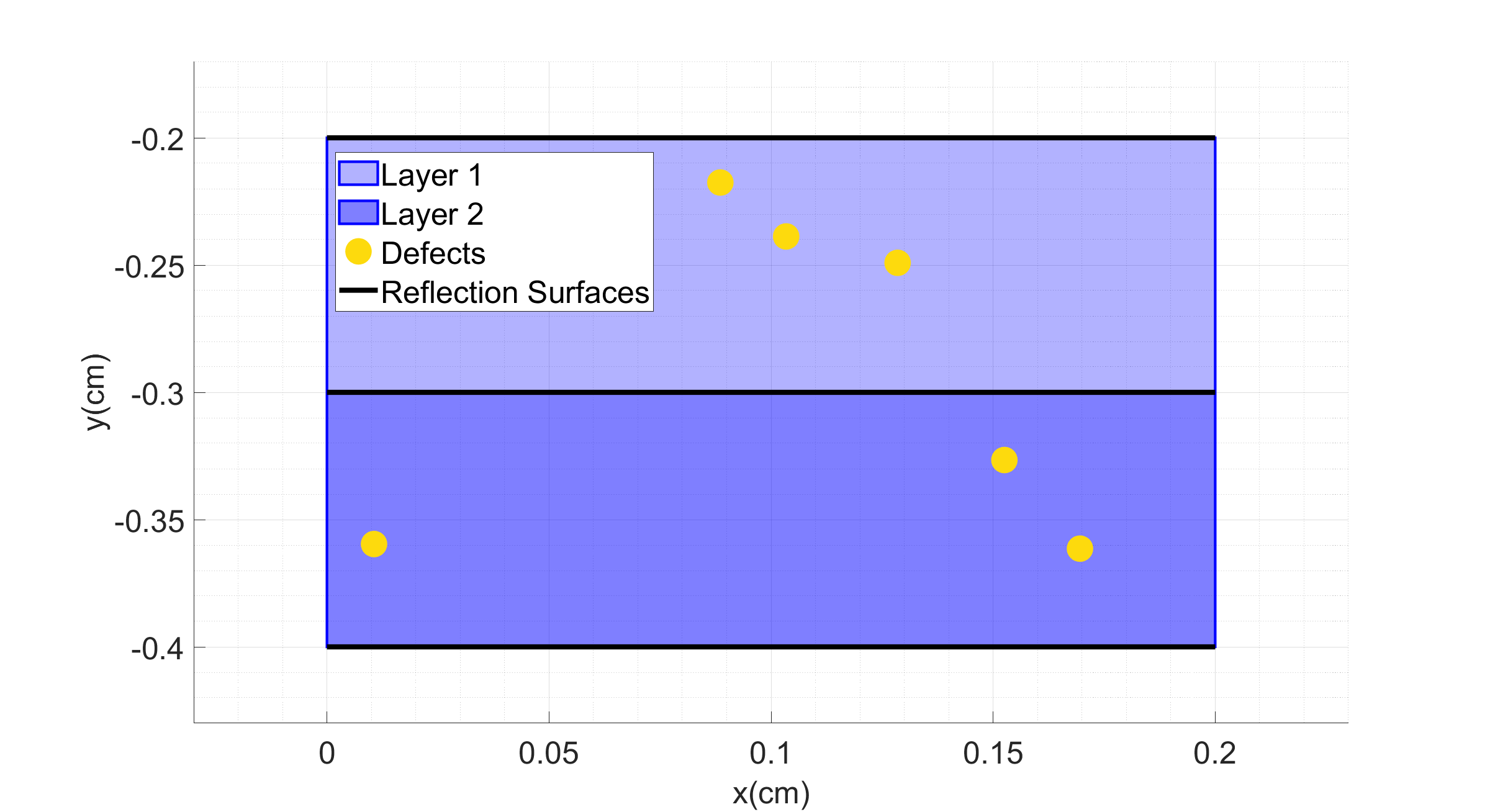}
    \caption{Illustration of the sample MUT with multiple defects}
    \label{fig:defects}
\end{figure}

In this section, we present the studies that we conducted for
the proposed system. In our simulations, we placed a sample material consisting of two layers, each containing six defects, within a 2D coordinate system, as shown in Fig. \ref{fig:defects}. 

\begin{figure}[t!]
    \centering
    \includegraphics[width=\linewidth]{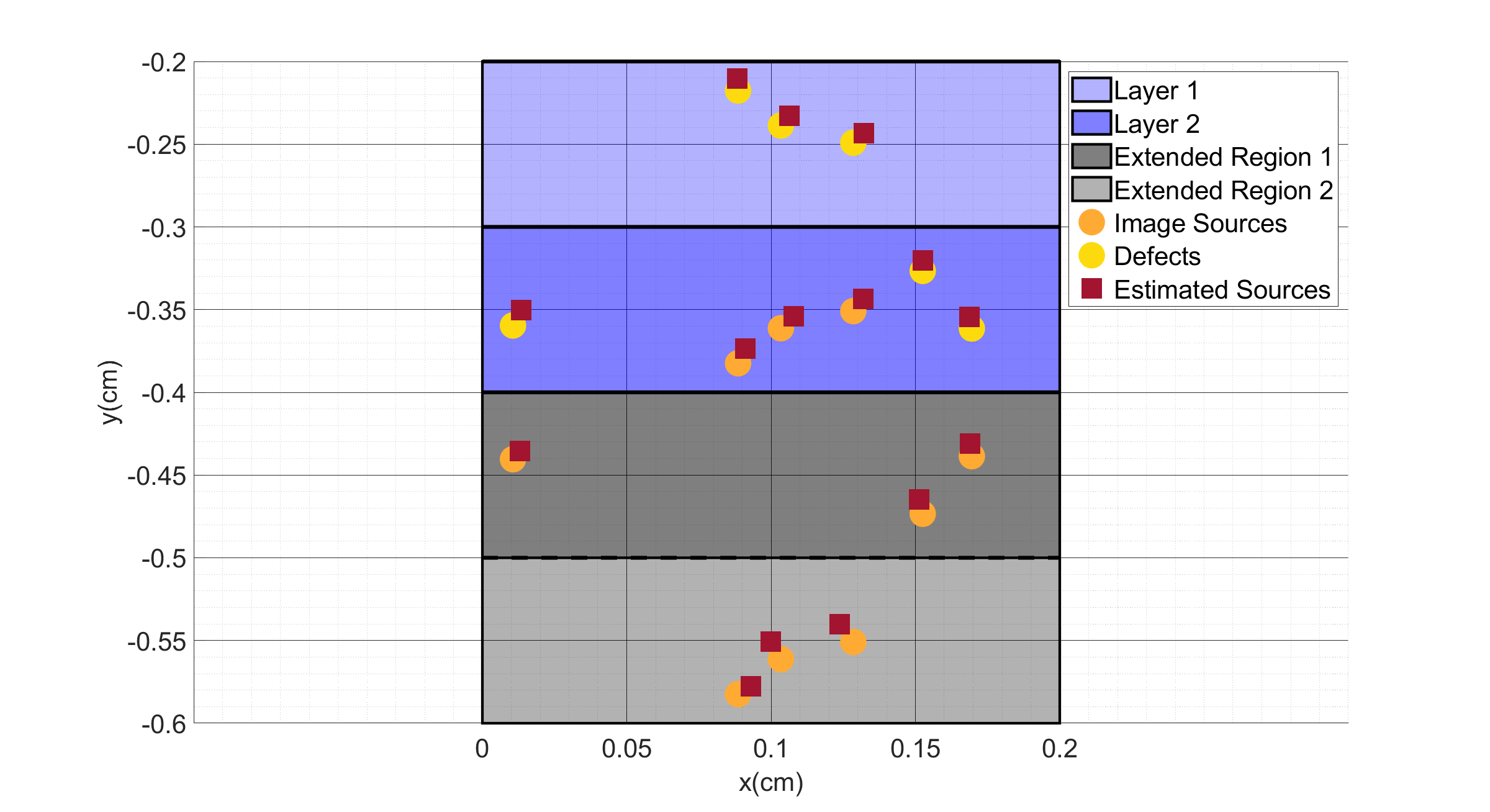}
    \caption{Estimated locations of the defects and the image sources on $2$D plane }
    \label{fig:imsources}
     \vspace{-4mm}
\end{figure}

\begin{figure}[t!]
    \centering
    \includegraphics[width=\linewidth]{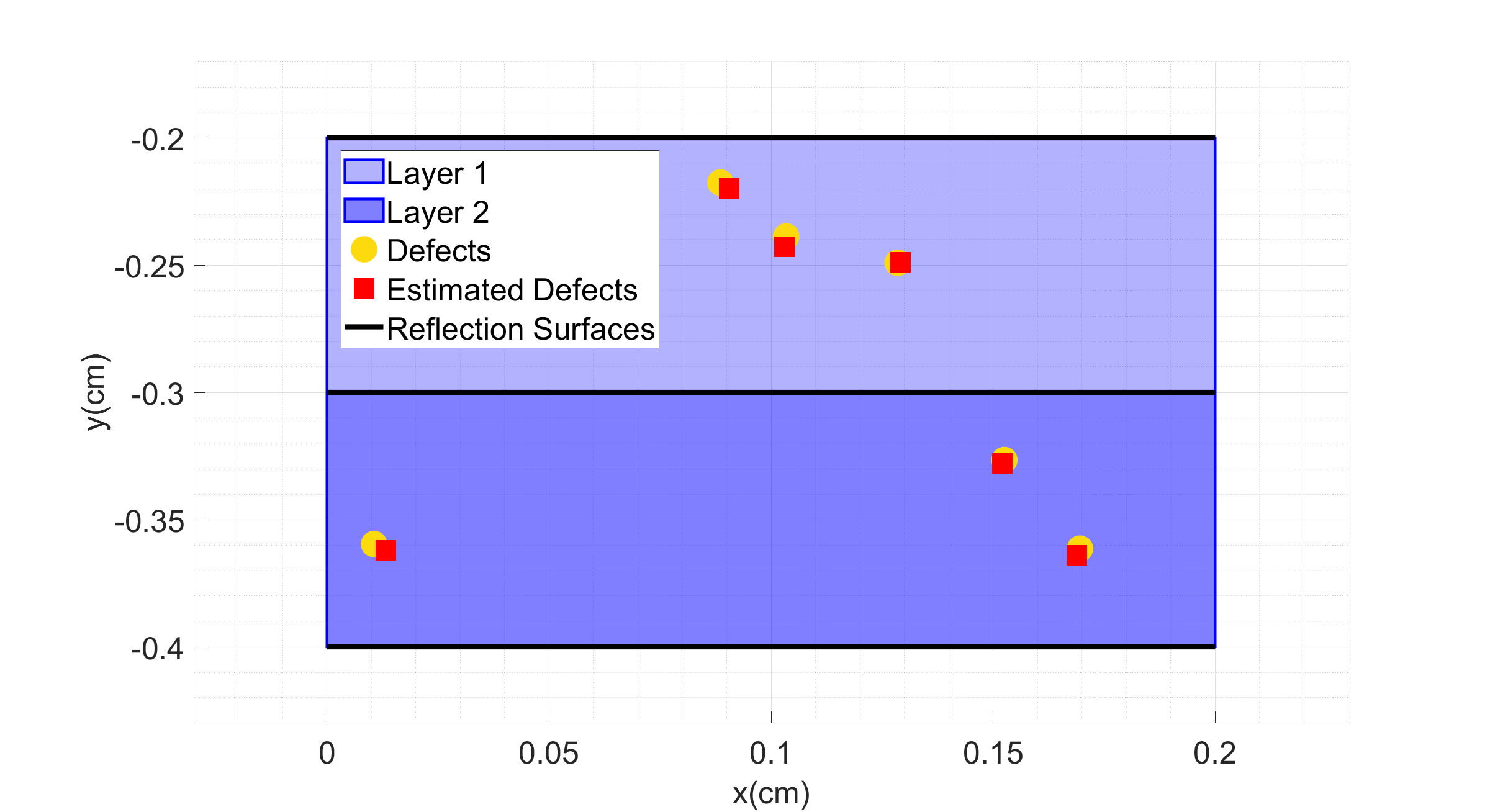}
    \caption{Comparison of existing defect locations with the estimated defect locations with defect reconstruction algorithm }
    \label{fig:estimdefects}
    \vspace{-4mm}
\end{figure}

The two layers are represented in different shades of blue for clarity and the defects are in yellow. Fig. \ref{fig:imsources} illustrates the estimated sources, including both defects and image sources, as determined through echo labeling. The defects and image sources are represented by yellow and orange circles, respectively, while the estimated sources are shown as red squares. The shaded regions can be considered as an extension of the material, where the image sources map onto 2D-coordinate system. From this figure, we can observe that echo labeling successfully detects all the sources. Furthermore, the estimated locations are close to the true positions, falling within the error bounds. This demonstrates the effectiveness of the echo-labeling method. Finally, Fig. \ref{fig:estimdefects} presents the reconstructed material using the Alg. \ref{DefRec}. From this figure, we can conclude that the EDM-based defect detection method performs effectively. The detected points are in close proximity to the actual defects, highlighting the success of the method in accurately identifying the defects. 

%% file: 5_conclusions.tex
In this study, we have shown that the multi-layer defect detection can be redefined as an echo labeling problem. The proposed method efficiently maps higher-order reflections to the correct defects by leveraging the rank properties of the EDM, demonstrating both speed and effectiveness. In future work, we aim to extend this approach to structures with arbitary medium-speeds and enhance the coverage of transmit antennas, thereby accelerating the object scanning process.

%% file: IEEE-conference.bbl
\begin{thebibliography}{10}
\providecommand{\url}[1]{#1}
\csname url@samestyle\endcsname
\providecommand{\newblock}{\relax}
\providecommand{\bibinfo}[2]{#2}
\providecommand{\BIBentrySTDinterwordspacing}{\spaceskip=0pt\relax}
\providecommand{\BIBentryALTinterwordstretchfactor}{4}
\providecommand{\BIBentryALTinterwordspacing}{\spaceskip=\fontdimen2\font plus
\BIBentryALTinterwordstretchfactor\fontdimen3\font minus
  \fontdimen4\font\relax}
\providecommand{\BIBforeignlanguage}[2]{{%
\expandafter\ifx\csname l@#1\endcsname\relax
\typeout{** WARNING: IEEEtran.bst: No hyphenation pattern has been}%
\typeout{** loaded for the language `#1'. Using the pattern for}%
\typeout{** the default language instead.}%
\else
\language=\csname l@#1\endcsname
\fi
#2}}
\providecommand{\BIBdecl}{\relax}
\BIBdecl

\bibitem{yakoleVoids}
E.~V. Yakovlev, K.~I. Zaytsev, N.~V. Chernomyrdin, A.~A. Gavdush, A.~K. Zotov,
  M.~Y. Nikonovich, and S.~O. Yurchenko, ``Non-destructive testing of composite
  materials using terahertz time-domain spectroscopy,'' in \emph{Optical
  Sensing and Detection IV}, vol. 9899.\hskip 1em plus 0.5em minus 0.4em\relax
  SPIE, 2016, pp. 201--207.

\bibitem{ospaldDelim}
F.~Ospald, W.~Zouaghi, R.~Beigang, C.~Matheis, J.~Jonuscheit, B.~Recur, J.-P.
  Guillet, P.~Mounaix, W.~Vleugels, P.~V. Bosom \emph{et~al.}, ``Aeronautics
  composite material inspection with a terahertz time-domain spectroscopy
  system,'' \emph{Optical Engineering}, vol.~53, no.~3, pp. 031\,208--031\,208,
  2014.

\bibitem{Lukas}
L.~B. Schäfer, U.~S. K.~P. Miriya~Thanthrige, and A.~Sezgin, ``Iteratively
  reweighted nuclear norm based distortion compensation for thz-tds,'' in
  \emph{Proc. IWMTS}, 2022, pp. 1--5.

\bibitem{TEOSingle}
S.~M. Teo, B.~K. Ofori-Okai, C.~A. Werley, and K.~A. Nelson, ``Invited article:
  Single-shot thz detection techniques optimized for multidimensional thz
  spectroscopy,'' \emph{Review of scientific instruments}, vol.~86, no.~5,
  2015.

\bibitem{mansourzadeh2021high}
S.~Mansourzadeh, D.~Damyanov, T.~Vogel, F.~Wulf, R.~B. Kohlhaas, B.~Globisch,
  T.~Schultze, M.~Hoffmann, J.~C. Balzer, and C.~J. Saraceno, ``High-power
  lensless thz imaging of hidden objects,'' \emph{IEEE Access}, vol.~9, pp.
  6268--6276, 2021.

\bibitem{Saraceno}
T.~Vogel, S.~Mansourzadeh, and C.~J. Saraceno, ``Single-cycle, 643 mw average
  power terahertz source based on tilted pulse front in lithium niobate,''
  \emph{Optics Letters}, vol.~49, no.~16, pp. 4517--4520, 2024.

\bibitem{acoustics}
I.~Dokmani{\'c}, R.~Parhizkar, A.~Walther, Y.~M. Lu, and M.~Vetterli,
  ``Acoustic echoes reveal room shape,'' \emph{Proc. of the National Academy of
  Sciences}, vol. 110, no.~30, pp. 12\,186--12\,191, 2013.

\bibitem{acusticGraph}
I.~Jager, R.~Heusdens, and N.~D. Gaubitch, ``Room geometry estimation from
  acoustic echoes using graph-based echo labeling,'' in \emph{Proc. ICASSP},
  2016, pp. 1--5.

\bibitem{Sampath}
U.~S. K.~P. Miriya~Thanthrige and A.~Sezgin, ``Tera-hertz sub-layer object
  identification using mca and dictionary learning,'' in \emph{Proc. IWMTS},
  2019, pp. 1--6.

\bibitem{SampathMIMO}
U.~S. K.~P. Miriya~Thanthrige, P.~Jung, and A.~Sezgin, ``Deep unfolding of
  iteratively reweighted admm for wireless rf sensing,'' \emph{Sensors},
  vol.~22, no.~8, 2022.

\bibitem{SampathML}
M.~Al-Askary, U.~S. Miriya~Thanthrige, P.~Pfeffer, L.~Wachter, G.~Schober, and
  A.~Sezgin, ``A comparative study of low-rank-plus-sparse matrix decomposition
  and machine learning for non-destructive air-ultrasound defect detection,''
  in \emph{Proc. EUSIPCO}, 2021, pp. 2109--2113.

\bibitem{Mohsen}
\BIBentryALTinterwordspacing
M.~Khalili, T.~Vogel, Y.~Wang, S.~Mansourzadeh, A.~Singh, S.~Winnerl, and C.~J.
  Saraceno, ``Microstructured large-area photoconductive terahertz emitters
  driven at high average power,'' \emph{Opt. Express}, vol.~32, no.~13, pp.
  22\,955--22\,969, Jun 2024. [Online]. Available:
  \url{https://opg.optica.org/oe/abstract.cfm?URI=oe-32-13-22955}
\BIBentrySTDinterwordspacing

\bibitem{DokmanicEDM}
I.~Dokmanic, R.~Parhizkar, J.~Ranieri, and M.~Vetterli, ``Euclidean distance
  matrices: essential theory, algorithms, and applications,'' \emph{IEEE Signal
  Processing Magazine}, vol.~32, no.~6, pp. 12--30, 2015.

\bibitem{SHABALIN201367}
A.~A. Shabalin and A.~B. Nobel, ``Reconstruction of a low-rank matrix in the
  presence of gaussian noise,'' \emph{Journal of Multivariate Analysis}, vol.
  118, pp. 67--76, 2013.

\bibitem{BronKerbosch}
C.~Bron and J.~Kerbosch, ``Algorithm 457: finding all cliques of an undirected
  graph,'' \emph{Communications of the ACM}, vol.~16, no.~9, pp. 575--577,
  1973.

\end{thebibliography}
